\documentclass[useAMS,usenatbib]{mn2e}
\usepackage{longtable}
\usepackage{graphicx}
\usepackage{lscape}
\usepackage{amsmath}
\usepackage{rotating}

\newcommand{\mnras}{MNRAS}
\newcommand{\aj}{AJ}
\newcommand{\apj}{ApJ}
\newcommand{\apjs}{ApJ}

\newcommand{\aap}{A\&A}

\newcommand{\pasp}{PASP}

\newcommand{\memsai}{Memorie della Societa Astronomica Italiana}

\newcommand{\Mwd}{\mbox{$M_\mathrm{wd}$}}

\newcommand{\Msun}{\mbox{$\mathrm{M}_{\odot}$}}

\newcommand{\Teff}{\mbox{$T_{\mathrm{eff}}$}}

\title[The  mass   function  of   DA  WDs]   {The  mass   function  of
  hydrogen-rich  white dwarfs:  robust  observational  evidence for  a
  distinctive high-mass excess near 1\,\Msun}

\author[A.  Rebassa-Mansergas et al.]{A.
  Rebassa-Mansergas$^{1}$\thanks{Email; arebassa@pku.edu.cn},
  M. Rybicka$^{2}$, X.-W.  Liu$^{1,3}$, Z. Han$^4$, E. Garc\'ia--Berro$^{5,6}$\\
$^{1}$  Kavli   Institute  for  Astronomy  and   Astrophysics,  Peking
  University, Beijing 100871, P.\,R.\,China\\
$^{2}$Copernicus Astronomical Center, Warszawa, Poland\\
$^{3}$ Department of Astronomy, Peking University, Beijing 100871,
P.\,R.\,China\\
$^{4}$ Key  Laboratory for  the Structure  and Evolution  of Celestial
Objects, Yunnan  observatories, Chinese Academy of  Sciences, P.O. Box
110, Kunming, \\650011, Yunnan Province, P.\,R.\,China\\
$^{5}$ Departament de F\'\i sica Aplicada, Universitat Polit\`ecnica
  de Catalunya, c/Esteve Terrades 5, 08860 Castelldefels,
  Spain\\ 
$^{6}$ Institute for Space Studies of Catalonia, c/Gran Capit\`a 2--4,
  Edif. Nexus 201, 08034 Barcelona, Spain
}

\begin{document}
\date{Accepted 2015. Received 2015; in original form 2015}
\pagerange{\pageref{firstpage}--\pageref{lastpage}} \pubyear{2015}
\maketitle

\begin{abstract}
The mass  function of hydrogen-rich  atmosphere white dwarfs  has been
frequently  found  to  reveal  a  distinctive  high-mass  excess  near
1\,\Msun.  However, a  significant excess of massive  white dwarfs has
not been  detected in  the mass  function of  the largest  white dwarf
catalogue to date from the Sloan Digital Sky Survey.  Hence, whether a
high-mass excess exists or not has remained an open question.  In this
work  we build  the  mass function  of the  latest  catalogue of  data
release 10  SDSS hydrogen-rich  white dwarfs,  including the  cool and
faint  population  (i.e.   effective temperatures  6,000  $\la$  \Teff
$\la$12,000\,K,  equivalent to  12 mag  $\la$ M$_\mathrm{bol}  \la 13$
mag).  We  show that the  high-mass excess  is clearly present  in our
mass function, and that it  disappears only if the hottest (brightest)
white dwarfs (those with  \Teff$\ga$12,000\,K, M$_\mathrm{bol} \la 12$
mag) are considered.   This naturally explains why  previous SDSS mass
functions failed at detecting a  significant excess of high-mass white
dwarfs.  Thus, our results provide additional and robust observational
evidence  for the  existence of  a distinctive  high-mass excess  near
1\,\Msun.  We investigate  possible origins of this  feature and argue
that the most  plausible scenario that may lead to  an observed excess
of massive  white dwarfs  is the  merger of the  degenerate core  of a
giant star with  a main sequence or a white  dwarf companion during or
shortly after a common envelope event.
\end{abstract}


\label{firstpage}

\section{Introduction}
\label{s-intro}

White dwarfs (WDs) are the typical  end products of most main sequence
stars (see \citealt{althausetal10-1} and references therein).  WDs are
therefore the most numerous stellar  remnants in the Galaxy.  The mass
distribution and  mass function  (MF) of  hydrogen-rich (DA)  WDs have
been  intensively  studied during  the  last  decades.  These  studies
reveal   a  clear   and  predominant   concentration  of   objects  at
$\sim$0.6\,\Msun\,   \citep[e.g.][]{koesteretal79-1,  holbergetal08-1,
  kepler13-1, kepleretal15-1}.  A  low-mass peak at $\sim$0.4\,\Msun\,
has been also frequently found \citep{liebertetal05-1, kepleretal07-1,
  kleinmanetal13-1}.  This  feature is believed  to arise as  a simple
consequence of binary star evolution.   In this scenario mass transfer
episodes truncate the evolution of the red giant exposing its low-mass
core,  which   later  becomes   a  (typically  He-core)   WD.   Strong
observational evidence in favour of  this hypothesis has been provided
during      the     last      few     years      \citep{marshetal95-1,
  rebassa-mansergasetal11-1,   kilicetal12-1},   although   some   few
low-mass WDs have  been observed that exhibit  neither radial velocity
variations, nor  infrared flux  excess, the  typical hallmarks  of WDs
with   close  companions   \citep{maxtedetal00-4,  napiwotzkietal07-1,
  kilicetal10-1}.

Recent observational studies  suggest also the existence  of an excess
of  massive  white   dwarfs  near  1\,\Msun\,  \citep{liebertetal05-1,
  giammicheleetal12-1,  rebassa-mansergasetal15-1}.   This feature  is
generally interpreted as the result  of WD+WD binary mergers. This, in
turn, may indicate that the merger  rate in the Galaxy is considerably
larger  than  expected.  Such  a  high  merger  rate may  have  strong
implications  in  the  production  of   type  Ia  supernovae  via  the
double-degenerate    channel     \citep{webbink84-1,    distefano10-1,
  jietal13-1}.  It  is also important  to keep in mind,  however, that
the WD MF that results from analysing the so far largest spectroscopic
sample    of    WDs    from    the   Sloan    Digital    Sky    Survey
\citep[SDSS;][]{ahnetal12-1}  displays no  evidence for  a distinctive
concentration   of   systems   at  those   specific   high-mass   bins
\citep{huetal07-1, kepleretal07-1, degennaroetal08-1}.  Whether or not
an excess of massive WDs exists is then a controversial issue.

In this  work we derive  the MF of  WDs from the  latest spectroscopic
catalogue of SDSS data release (DR) 10 \citep{kepleretal15-1} and show
that a clear  excess near $\sim$1\,\Msun\, can only be  seen if WDs of
bolometric  magnitude fainter  than $\sim$12  mag are  considered.  We
interpret  this result  as  a robust  observational  evidence for  the
existence of an excess of massive  WDs in the Galaxy.  We also discuss
possible origins leading to this feature.

\section{SDSS DR10 WDs}

In this section we introduce the DA WD sample studied in this work and
provide details on the space density and mass determinations that lead
us to derive the MF.

\subsection{Masses}

Our sample of study is the latest version of the SDSS WD spectroscopic
catalogue  \citep{kepleretal15-1},  which   currently  contains  about
30\,000 objects.  Given that we are  interested in analysing the MF of
DA WDs,  we select only those  of this subtype.  For  these, effective
temperatures  and   surface  gravities  have  been   provided  by  the
sub-sequent   SDSS   WD  catalogues   during   the   last  ten   years
\citep{kleinmanetal04-1,      eisensteinetal06-1,      girvenetal11-1,
kleinmanetal13-1,  kepleretal15-1}. These  have been  obtained fitting
the Balmer  lines sampled  by the SDSS  spectra with  model atmosphere
spectra \citep[e.g.][]{koester10-1}.   Given that each SDSS  DR yields
re-calibrated  spectra  respect  to  previous releases,  each  of  the
aforementioned SDSS WD catalogues provides updated values of effective
temperatures and  surface gravities obtained by  re-analysing the SDSS
re-calibrated spectra.   In this work  we adopt the most  updated (and
therefore, more accurate) available values.

\begin{figure}
\begin{center}
\includegraphics[angle=-90,width=\columnwidth] {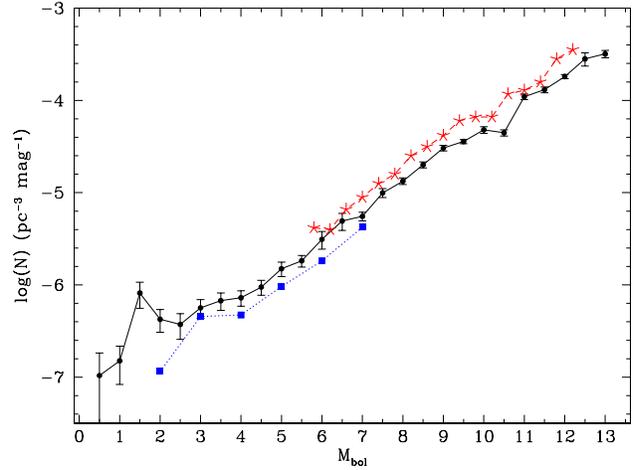}
\caption{\label{f-lumfunc} DA WD LF derived  in this work (black solid
  dots).    For   comparison   we   also  show   the   SDSS   LFs   of
  \citet{degennaroetal08-1}  (red  stars)  and  \citet{torresetal14-1}
  (blue squares).}
\end{center}
\end{figure}

It is important to emphasize that  fitting the Balmer lines with model
atmosphere models results in over-estimated surface gravity values for
DA WDs cooler than $\sim$13\,000\,K \citep{koesteretal09-1}, a problem
related to the 1D treatment  of convective energy transport within the
framework  of   the  mixing-length   theory  \citep{tremblayetal11-1}.
However, 3D model corrections are  now available, which can be applied
to  correct  for  this  problem  \citep{tremblayetal13-1}.   We  hence
applied these corrections to all cool WDs and thus re-derived reliable
effective temperatures and surface gravities.

Finally,  using  the  cooling   tracks  of  \citet{renedoetal10-1}  we
obtained  the   masses,  the   bolometric  magnitudes  and   the  SDSS
$u,g,r,i,z$  absolute   magnitudes  of  all   our  DA  WDs   from  the
temperatures   and   surface   gravities   measured   observationally.
Distances  were  calculated  from  the distance  moduli,  taking  into
account the  apparent SDSS  $u,g,r,i,z$ magnitudes.   For the  sake of
comparison, we repeated this exercise using the updated cooling models
of
\citet{bergeronetal95-2}\footnote{http://www.astro.umontreal.ca/$\sim$bergeron/CoolingModels/}.
We  found no  substantial  difference between  the masses,  bolometric
magnitudes  and  distances  derived  from the  two  cooling  sequences
considered.  Extinction corrections were also found to be negligible.

\begin{table}
\caption{\label{t-LF}    The   values    of    the    LF   shown    in
  Figure\,\ref{f-lumfunc}  per  M$_\mathrm{bol}$  bin. The  units  are
  pc$^{-3}$ mag$^{-1}$.}  \setlength{\tabcolsep}{1.1ex}
\begin{small}
\begin{center}
\begin{tabular}{cccccccc}
\hline
\hline
 M$_\mathrm{bol}$ & $\log$(N) & M$_\mathrm{bol}$ & $\log$(N) & M$_\mathrm{bol}$ & $\log$(N) & M$_\mathrm{bol}$ & $\log$(N)\\
\hline
0.5  &    $-6.98$  & 4.0  &    $-6.14$  &  7.5 & $-5.00$ & 10.5 & $-4.35$ \\
1.0  &    $-6.82$  & 4.5  &    $-6.02$  &  8.0 & $-4.87$ & 11.0 & $-3.96$ \\
1.5  &    $-6.09$  & 5.0  &    $-5.82$  &  8.5 & $-4.70$ & 11.5 & $-3.88$ \\
2.0  &    $-6.37$  & 5.5  &    $-5.74$  &  9.0 & $-4.52$ & 12.0 & $-3.74$ \\
2.5  &    $-6.43$  & 6.0  &    $-5.51$  &  9.5 & $-4.45$ & 12.5 & $-3.55$ \\
3.0  &    $-6.25$  & 6.5  &    $-5.31$  & 10.0 & $-4.32$ & 13.0 & $-3.50$ \\
3.5  &    $-6.17$  & 7.0  &    $-5.26$  &      &         &      &         \\
\hline   
\end{tabular} 
\end{center}  
\end{small}     
\end{table}   

\subsection{Space densities}

The 1$/V_{\rm max}$ method  \citep{schmidt68-1, green80-1} is a widely
used  technique  to  derive  the space  densities  of  WDs  \citep[see
  e.g.][]{liebertetal05-1,         harrisetal06-1,         huetal07-1,
  degennaroetal08-1,                       rebassa-mansergasetal15-1}.
\citet{geijoetal06-1} demonstrated that the 1$/V_{\rm max}$ method not
only is a  good estimator of the  WD space densities but  also that it
provides a reliable characterization of the WD luminosity function. In
this paper  we also  adopt the  1$/V_{\rm max}$  method to  derive the
space  density of  DA WDs  in the  SDSS.  That  is, we  calculated the
maximum volume  $V_\mathrm{WD}$ in  which each of  our WDs  would have
been  detected  given   the  magnitude  limits  of   the  SDSS  survey
\citep{rebassa-mansergasetal15-1}:

\begin{eqnarray}
\label{eq-vol}
V_\mathrm{WD} = V_\mathrm{max}-V_\mathrm{min} = {\sum_{i=1}^{n_{\rm plate}}}\,\frac{\omega_i}{4\pi}\int_{d_\mathrm{min}}^{d_\mathrm{max}}e^{-z/z_0}~4{\pi}r^2dr = \nonumber\\
= -\sum_{i=1}^{n_{\rm plate}} \frac{z_0 \times \omega_i}{\left |\sin{b}\right |}\left [ \left (r^2+\frac{2z_0}{\left |\sin{b}\right |}r+\frac{2z_0^2}{\left |\sin{b}\right |^2}  \right )\! e^{-\frac{r\left |\sin{b}\right |}{z_0}} \right ]_{d_\mathrm{min}}^{d_\mathrm{max}} 
\end{eqnarray}

\noindent where $b$ is the Galactic latitude of the considered WD, and
$\omega_i$  is the  solid angle  in  steradians covered  by each  SDSS
plate\footnote{Each  SDSS  plate   covers  an  area  of   the  sky  of
  $\sim$7\,deg$^2$.  Since the number of SDSS DR\,10 plates is 4\,171,
  the  total  area  is approximately  $\sim$29\,000\,deg$^2$,  without
  taking into account the  overlapping areas.}.  The factor $e^{-z/z_0
}$ takes  into account  the non-uniform distribution  of stars  in the
direction  perpendicular  to  the  Galactic disc,  where  $z=r  \times
\sin(b)$ is the  distance of the WD from the  Galactic plane and $z_0$
is   the   scale   height,   which    is   assumed   to   be   250\,pc
\citep{liebertetal05-1,  huetal07-1}.   We  considered the  lower  and
upper  $g$ magnitude  limits  of  each SDSS  plate,  which define  the
minimum  and maximum  volumes,  $V_\mathrm{min}$ and  $V_\mathrm{max}$
(where   $V_\mathrm{WD}   =  V_\mathrm{max}-V_\mathrm{min}$).    These
magnitude  limits   corresponded  to  the  minimum   and  maximum  $g$
magnitudes  among all  spectroscopic  sources observed  by each  plate
\citep{rebassa-mansergasetal15-1}.   However, as  we  show below,  our
sample  is restricted  to WDs  of $g  \leq$\,19\,mag.  Hence,  all the
upper magnitude  limits above  this value  were set  to $g=$\,19\,mag.
Moreover, in  the cases  where two  or more  plates observed  the same
region of sky,  we considered the overlapping region  with the largest
volume.

It has to be noted however that the overall spectroscopic completeness
of  SDSS WDs  is $\sim$40  per  cent, and  that  it is  found to  vary
significantly   in    colour   space   \citep{gentilefusilloetal15-1}.
Therefore, a  spectroscopic completeness correction needs  to be taken
into     account    in     our     space    density     determinations
\citep[e.g.][]{degennaroetal08-1}.    We   did    this   as   follows.
\citet{gentilefusilloetal15-1} provide  a list of  23\,696 photometric
sources with available proper motions and $g\leq$ 19 mag from the full
DR\,10 footprint with a high-confidence probability for being a WD, of
which 5\,857 have available SDSS spectra.   For each of the 5\,857 WDs
in the  spectroscopic list  we obtained  their $u-g$,  $g-r$,$r-i$ and
$i-z$ colours, and defined a four-dimension (one dimension per colour)
sphere   of   0.05   colour   radius  around   each   of   them   (see
\citealt{camachoetal14-1} for further details).  Within each sphere we
obtained  the   number  of  photometric  ($N_\mathrm{phot   WD}$)  and
spectroscopic ($N_\mathrm{spec WD}$) WDs  among the 23\,696 and 5\,857
lists respectively and calculated the spectroscopic completeness as $C
= N_\mathrm{spec  WD}/N_\mathrm{phot WD}$.  The space  density of each
WD,  $1/V_\mathrm{WD}$,  was  then   corrected  by  the  spectroscopic
completeness via $1/V_\mathrm{WD} \times 1/C$.

\begin{figure}
\begin{center}
\includegraphics[angle=-90,width=\columnwidth] {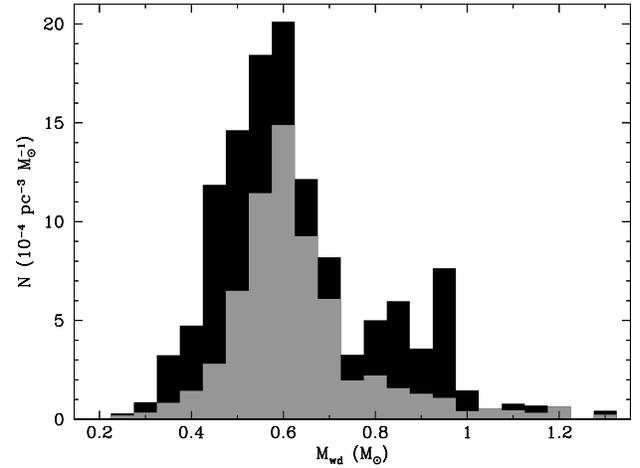}
\caption{\label{f-mfunc} MF of DA WDs  derived in this work (black for
  WDs     of    M$_\mathrm{bol}\leq13$,     gray     for    WDs     of
  M$_\mathrm{bol}\leq12$).   The  high-mass   excess  near  1\,\Msun\,
  disappears when considering DA WDs with M$_\mathrm{bol}\leq12$.}
\end{center}
\end{figure}

Of course, the above exercise limited our sample of study to 5\,857 DA
WDs, namely those with available probabilities for being a WD based on
their  photometric  colours  and  proper  motions,  as  well  as  with
available  values  of  effective temperatures  and  surface  gravities
obtained fitting their SDSS spectra.   This excluded, for example, all
confirmed   spectroscopic  WDs   with  g$>$19\,mag.    Therefore,  the
possibility  exists that  our  considered sample  cannot  be taken  as
complete  (this completeness  is  not  the spectroscopic  completeness
above discussed).  A  fairly standard way to check this  is to compute
the   average   value    $\langle   V-V_\mathrm{min}\rangle/   \langle
V_\mathrm{max}-V_\mathrm{min} \rangle$ (where $V$ is the volume of the
WD),  which   should  be  $\sim   0.5$  if  the  sample   is  complete
\citep{green80-1}.   In   our  case   the  sample   of  DA   WDs  with
spectroscopic determinations resulted in an average value of 0.48.  We
can therefore consider this sample as reasonably complete.

Our selected DA  WD sample has a mean mass  of 0.59 $\pm$ 0.12\,\Msun,
in   agreement   with   those   found   in   previous   SDSS   studies
\citep[e.g.][]{kepleretal07-1,         degennaroetal08-1}.         The
signal-to-noise ratio of  the SDSS spectra is $>$10 in  95 per cent of
the  cases, which  leads  to average  uncertainties  in the  effective
temperature    and   mass    determinations   of    $\sim$300\,K   and
$\sim$0.03\,\Msun,  respectively.  The  mass uncertainties  are  below
0.1\,\Msun\, in 99.5 per cent of the cases.

The WD luminosity  function (LF) obtained from our  selected sample is
shown in  Figure\,\ref{f-lumfunc} (see also  Table\,\ref{t-LF}), where
the   uncertainties   are   calculated   following   \citet{boyle89-1}.
Inspection of this  figure reveals that our LF covers  a wide range of
M$_\mathrm{bol}$ bins, reaching values as  large as 0.5 mag.  However,
we note that the shape of the LF should be taken with some caution for
M$_\mathrm{bol}\leq  2.5$ due  to the  increasing uncertainties.   For
comparative  purposes,  in Fig.~\ref{f-lumfunc}  the  SDSS  WD LFs  of
\citet{degennaroetal08-1}   and    \citet{torresetal14-1}   are   also
displayed.  It is clear that the three works yield LFs of very similar
shapes, although the absolute levels  are somewhat different.  This is
most likely a consequence of  the slightly different completenesses of
the  three samples.   In particular,  samples that  are more  complete
result in  larger space densities.   The similarity between  the three
LFs indicates  that (1)  our spectroscopic completeness  and $1/V_{\rm
  max}$  corrections are  properly done;  and that  (2) our  sample is
indeed  reasonably  complete  (at  least within  the  context  of  the
magnitude limits of SDSS, i.e.  it does not account for populations of
WDs that are too faint (M$_\mathrm{bol}>13$ mag) to be included in the
observed sample).

\section{The DA WD mass function}

\begin{table}
\caption{\label{t-frac}    Fraction   of    DA   WDs    fainter   than
  M$_\mathrm{bol}    =   12$\,mag    as   a    function   of    mass.}
\setlength{\tabcolsep}{0.6ex}
\begin{small}
\begin{center}
\begin{tabular}{lcccccc}
\hline
\hline
Mass (\Msun) & 0.55-0.65 & 0.65-0.75 & 0.75-0.85 & 0.85-0.95  &0.95-1.05 & $>$1.05 \\
Fraction &  0.06 & 0.07 & 0.18 & 0.16  & 0.21 & 0.15 \\
\hline      
\end{tabular} 
\end{center}  
\end{small}  
\end{table}

The  MF   of  our  sample   of  5\,857   DA  WDs  is   illustrated  in
Fig.\,\ref{f-mfunc}.     It   displays    a   predominant    peak   at
$\sim$0.6\,\Msun\,,  a  classical  feature that  has  been  previously
identified   in   numerous   studies   \citep[e.g.][]{koesteretal79-1,
  holbergetal08-1, kepleretal07-1, degennaroetal08-1}.  Our MF reveals
also  the  existence of  low-mass  WDs  (\Mwd$<0.55$\,\Msun) that  are
thought  to  arise as  a  consequence  of  mass transfer  in  binaries
\citep{marshetal95-1,  rebassa-mansergasetal11-1}.  Low-mass  WDs have
been     also     frequently     identified    in     previous     MFs
\citep[e.g.][]{liebertetal05-1,    kepleretal07-1}.      Finally,    a
distinctive high-mass  excess can be seen  in the $\sim$0.8-1\,\Msun\,
mass interval of our MF.  This feature  has been found not only in MFs
derived     from    magnitude-limited     samples     of    DA     WDs
\citep{liebertetal05-1,  rebassa-mansergasetal15-1}, but  also in  the
mass distribution of local (and therefore volume-limited sample of) DA
WDs  \citep{giammicheleetal12-1}.  However,  it has  to be  emphasized
that previous  MFs obtained from  earlier releases  of the SDSS  DA WD
catalogue do not display a  significant excess of massive white dwarfs
at  those   specific  mass  bins   \citep{kepleretal07-1,  huetal07-1,
  degennaroetal08-1}.   We   investigate  this  discrepancy   in  what
follows.

The main difference  between the analysis presented here  and those of
\citet{kepleretal07-1, huetal07-1, degennaroetal08-1}  is that, unlike
us,  they only  considered  DA  WDs with  \Teff$\ga$12\,000-13\,000\,K
(i.e.  WDs  of M$_\mathrm{bol}\la$12). The  reason for this  is simply
because   3D  model   atmosphere   corrections  for   such  cool   WDs
\citep{tremblayetal13-1}  were  not available  at  that  time.  It  is
therefore possible that massive DA WDs were systematically excluded as
a  consequence of  only considering  WDs of  M$_\mathrm{bol}$ brighter
than $\sim$12 mag.  Indeed, and as can be seen in Fig.\,\ref{f-mfunc},
the  excess of  massive WDs  disappears if  we apply  this cut  to our
sample.

It is well known that massive WDs ($\ga$0.8\,$\mathrm{M}_{\sun}$) cool
down  considerably  slower  than  typical  WDs  of  $\sim$0.6\,\Msun\,
\citep{Althauseta05-1, Althausetal07-1}.   However, the  main sequence
progenitor lifetimes  (for solar  metallicities) are much  shorter for
the     precursors     of     massive    WDs     --     see,     e.g.,
\cite{pietrinfernietal04-1}.  This implies that, roughly speaking, for
decreasing   luminosities  the   fraction  of   massive  WDs   becomes
increasingly larger,  simply because these  WDs have had more  time to
cool down  than regular ones.  However,  we note that the  validity of
this assessment depends on the  specific star formation history of the
Galactic  disc, as  well as  on  possible metallicity  effects on  the
initial-to-final  mass  relation  and   on  the  white  dwarf  cooling
\citep[e.g.][]{romeroetal15-1, althausetal15-1}.  Hence, we checked if
the fraction of  massive WDs in our sample  increases considerably for
increasing bolometric magnitudes, and we found that this is indeed the
case  (see  Table\,\ref{t-frac}).   This naturally  explains  why  the
fraction  of massive  WDs excluded  in previous  analyses of  the SDSS
sample is larger when  the M$_\mathrm{bol}\leq12$~mag magnitude cut is
applied to the observed sample.

In the following  Section we argue the most  plausible explanation for
the existence of an overabundance of  massive WDs may be the merger of
the degenerate core of a giant  star with its companion (either a main
sequence  star or  a WD)  during or  shortly after  a common  envelope
event.  Hence, we expect a large  fraction of our observed massive WDs
to be the result  of such mergers.  In such cases  the lifetime of the
binary system  before the  merger takes  place should  be of  the same
order of the progenitor lifetimes of single massive WDs, otherwise WDs
that result from  such mergers would not have had  enough time to cool
down  and   would  therefore  not   be  excluded  when   applying  the
M$_\mathrm{bol}\leq12$~mag  cut.  The  timescale  for  a merger  event
during   common  envelope   can   be  as   short  as   $\sim$0.15\,Gyr
\citep{briggsetal15-1}, which  is considerably  faster than  e.g.  the
0.25\,Gyr needed for  the progenitor of a 0.8\,\Msun\, WD  to become a
WD    \citep{pietrinfernietal04-1}.    Thus,    concluding   that    a
M$_\mathrm{bol}\leq12$ cut  equally excludes massive WDs  that evolved
as single stars and massive WDs that  may arise as a result of mergers
seems to be a reasonable assumption.

\section{On the possible origin of the high-mass excess}

In the previous  section we provided further  and robust observational
evidence  for the  existence  of  an excess  of  massive  DA WDs  near
1\,\Msun.  In this section we  briefly discuss the possible origins of
this feature.

\begin{itemize}

\item The first possible explanation is that a large number of massive
  DA  WDs  are  formed  by   single  star  evolution.   For  instance,
  \citet{ferrarioetal05-1}   proposed  an   specific  shape   for  the
  initial-to-final mass  relationship that  accounts for  $\sim$28 per
  cent of all single WDs  having masses in excess of $\sim$0.8\,\Msun.
  However,  detailed numerical  simulations have  recently shown  that
  such an  initial-to-final mass  relationship cannot account  for the
  existence of all massive WDs \citep{rebassa-mansergasetal15-1}.

\item A  recent burst in  the star  formation history of  the Galactic
  disc may  contribute to  increase the  fraction of  observed massive
  WDs.  Indeed,  \cite{rowell13-1} suggested  that the  star formation
  rate (SFR)  has two  broad peaks  at around 2  and 7\,Gyr  ago. This
  could increase the number of massive WDs during the last 2 Gyr given
  their     shorter     main     sequence     lifetimes.      However,
  \citet{rebassa-mansergasetal15-1} also demonstrated  that such a SFR
  cannot explain the observed overabundance of massive WDs.

\item A  third explanation for  the existence  of a large  fraction of
  massive  WDs could  be  that the  1/$V_\mathrm{max}$ method  somehow
  overestimates  the space  density of  massive WDs.   Since such  WDs
  cannot be  seen out to  large distances,  they are assigned  a large
  weight by the 1/$V_\mathrm{max}$  method, hence largely contributing
  to the space  density and MF. However, the  volume-limited sample of
  DA WDs,  which does  not require the  1/$V_\mathrm{max}$ correction,
  displays also a clear excess of massive WDs near 1\,\Msun.

\item  The stellar  parameters derived  from the  spectra of  cool WDs
  required   3D   corrections  \citep{tremblayetal13-1}.    If   these
  corrections are  somehow inaccurate,  then we  would expect  a clear
  dependence of the  mass on the effective temperature.   We find that
  this is not  the case in our sample (see  Fig.\,\ref{f-mt}).  We can
  safely conclude then that 3D corrections are not responsible for the
  observed overabundance of massive WDs.

\item The  typical SDSS exposure  times are 30--90\,min,  depending on
  the considered  plate.  This  implies that  massive WDs  observed by
  SDSS may be associated  to intrinsically lower signal-to-noise ratio
  spectra,  as these  WDs  are smaller  and  therefore less  luminous.
  Hence,  the mass  uncertainties might  be systematically  higher for
  massive WDs.   The mass uncertainties  of our considered  sample are
  below  0.1\,\Msun\,  in  99.5  per  cent  of  the  cases.   We  thus
  re-derived the MF excluding all  WDs with errors above 0.03\,\Msun\,
  and  found that  the MF  that results  from this  exercise does  not
  differ  significantly   from  the   one  obtained  using   the  full
  sample. Therefore, the mass uncertainties  are unlikely to be at the
  root of the observed excess of massive WDs.

\begin{figure}
\begin{center}
\includegraphics[angle=-90,width=\columnwidth] {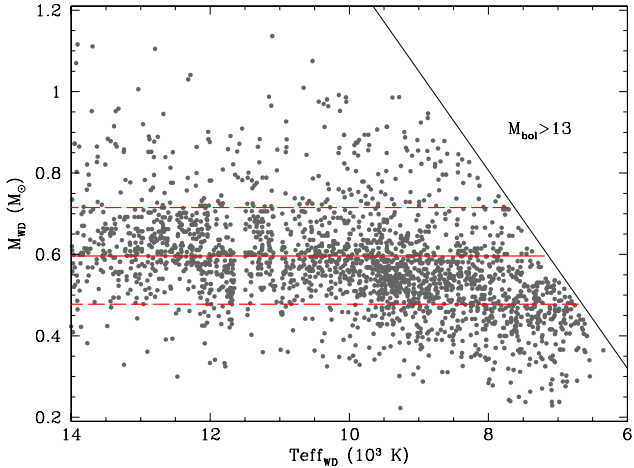}
\caption{\label{f-mt}  DA   WD  mass   as  a  function   of  effective
  temperature.   The  red  solid   line  indicates  the  average  mass
  $<$\Mwd$>$ of  the sample, the  red dashed lines  indicate $<$\Mwd$>
  \pm\, \sigma$.   The area delimited by  the solid black line  is for
  M$_\mathrm{bol} >  13$ mag  and it is  empty due to  our cut  in the
  luminosity function (see Fig.\,\ref{f-lumfunc}).}
\end{center}
\end{figure}

\item The  merger of  two WDs  is currently  the most  widely accepted
  explanation  for  the   existence  of  an  excess   of  massive  WDs
  \citep{marshetal97-1,          vennes99-1,          liebertetal05-1,
    giammicheleetal12-1, rebassa-mansergasetal15-1}.   If that  is the
  case, then  a large fraction  of all massive  WDs is expected  to be
  magnetic   \citep{garcia-berroetal12-1,   jietal13-1}.    Population
  synthesis studies however do not predict more than $\sim$10 per cent
  of the entire WD population being the result of WD+WD binary mergers
  \citep[e.g.][]{hanetal94-1,         han98-1,         toonenetal12-1,
    garcia-berroetal12-1}.    Although  these   simulations  generally
  assume  a constant  SFR,  the  results are  not  expected to  change
  considerably  when assuming  e.g.  a  bimodal  SFR such  as the  one
  suggested by \citet{rowell13-1}.  Therefore, if the high-mass excess
  arises as  a consequence of WD+WD  mergers, then the merger  rate in
  the Galaxy should be much  higher than currently expected.  This has
  strong implications for the production  of type Ia supernovae within
  the double  degenerate scenario \citep{distefano10-1}.  For  that to
  be the case, the merger of the two WDs must exceed the Chandrasekhar
  mass limit of $\sim$1.4\,\Msun, which  seems to be unlikely based on
  our (and  also previously  published) MF,  as it  does not  reveal a
  significant excess of WDs more massive than 1\,\Msun. However, it is
  important  to  mention  that   WDs  of  mass  $\ga$\,1\,\Msun\,  and
  effective   temperatures   below   10\,000\,K  are   associated   to
  M$_\mathrm{bol}  \ga  13$  mag \citep{renedoetal10-1},  hence  these
  massive   WDs    are   under-represented   in   our    sample   (see
  Fig.\,\ref{f-lumfunc}).

\item An additional explanation leading  to the formation of high-mass
  ($\ga$0.8\,\Msun)  WDs is  the merger  of the  degenerate core  of a
  giant or  asymptotic giant branch star  with a WD companion  after a
  common envelope  event \citep{kashi+soker11-1}.  In this  scenario a
  circumbinary disc is formed around  the two stars from material that
  remains bound  after the common  envelope phase. The  interaction of
  the circumbinary  disc with  the binary  system reduces  the orbital
  separation and results in the merger  of the (still hot) core of the
  giant and  the WD companion.  Whilst  this so-called core-degenerate
  scenario  is proposed  as a  viable channel  for type  Ia supernovae
  \citep{ilkov+soker12-1,        ilkov+soker13-1,       sokeretal13-1,
  sokeretal14-1, soker15-1}, the resulting mergers which do not exceed
  the   Chandrasekhar    mass   will   form   massive    WDs   instead
  \citep{aznar-siguanetal15-1}. In a similar way, massive WDs may form
  as a  result of  the merger  of the  degenerate core  of a  giant or
  asymptotic giant star with a main sequence companion during a common
  envelope  phase  \citep{briggsetal15-1}.    After  the  envelope  is
  ejected  the resulting  merger  will evolve  through the  asymptotic
  giant branch in the same was as a single star and form a massive WD.
\end{itemize}

Based on the  previous discussion we consider that  the most plausible
scenarios that may lead to the  observed excess of massive WDs involve
the merger  of the  degenerate cores  of giant  stars with  their main
sequence/WD companions during/after the  common envelope phase, and/or
the merger of WD+WD binaries. However, \citet{briggsetal15-1} suggests
that   WD+WD  mergers   typically  produce   WDs  more   massive  than
1\,\Msun.  If is  the  case,  then the  WD+WD  merger  channel is  not
expected to significantly  contribute to an excess of  massive WDs, as
our  MF  clearly  reveals  a  scarcity  of  systems  above  1\,\Msun\,
(Figure\,\ref{f-mfunc}). Furthermore,  the number  of WD  mergers that
form  through the  core-degenerate channels  is predicted  to be  much
larger than  the number of WDs  that result from the  merging of WD+WD
binaries  \citep[see   e.g.][]{garcia-berroetal12-1,  briggsetal15-1}.
Finally, it has to be  emphasized that the predicted mass distribution
of core-degenerate mergers  peaks at $\sim$0.8-0.9\,\Msun\, (depending
on the value of common envelope efficiency assumed in the simulations)
and   smoothly    declines   towards    lower   and    larger   values
\citep{briggsetal15-1}.   The expected  population of  core-degenerate
mergers thus falls precisely within the mass range where the high-mass
excess is observed in our MF (Figure\,\ref{f-mfunc}).

\section{Summary and conclusions}

We have  obtained the MF of  the latest catalogue of  SDSS (DR\,10) DA
WDs, including  for the  first time  the cool  and faint  (i.e.  6,000
$\la$ \Teff $\la$12,000\,K, 12 mag  $\la$ M$_\mathrm{bol} \la 13$ mag)
population.  We demonstrate  that a clear high-mass  excess is present
in  our  MF, which  disappears  if  only hot  and  bright  DA WDs  are
considered  (\Teff$\ga$12,000\,K, M$_\mathrm{bol}  \la  12$ mag).   We
interpret  our  result  as  an  additional  and  robust  observational
evidence  for  the existence  of  a  high-mass excess  near  1\,\Msun.
Although  the  merger  of  WD+WD  binaries  appears  as  a  reasonable
explanation  of   this  observed  feature,   sophisticated  population
synthesis  studies  have  shown   this  channel  does  not  contribute
significantly  to  explain  the   observed  excess  of  massive  white
dwarfs. Thus,  we argue  that the most  plausible scenario  leading to
this  feature is  the merger  of  the degenerate  core of  a giant  or
asymptotic  giant branch  star with  a main  sequence or  WD companion
during, or shortly after, a common envelope episode.

\section*{Acknowledgments}
ARM  acknowledges  financial  support from  the  Postdoctoral  Science
Foundation of China  (grants 2013M530470 and 2014T70010)  and from the
Research  Fund  for International  Young  Scientists  by the  National
Natural  Science   Foundation  of  China  (grant   11350110496).   XWL
acknowledges support  by the  National Key  Basic Research  Program of
China  2014CB845700.   ZH  acknowledges  financial  support  from  the
National Natural  Science Foundation of China  (grant 11390374).  EG-B
was partially supported by MCINN  grant AYA2011-23102, by the European
Union FEDER funds, and by the AGAUR.

\end{document}